\documentclass[twocolumn,preprintnumbers,msmath,amssymb,prl,floatfix,final]{revtex4-2}
\usepackage{graphicx}
\usepackage{dcolumn}
\usepackage{float}
\usepackage{bm}
\usepackage{units}
\usepackage{amsmath}
\usepackage{mciteplus}
\usepackage[utf8]{inputenc}
\makeatletter 
\def\@citess#1{\textsuperscript{[#1]}}
\makeatother 
\usepackage{comment}
\usepackage{xcolor, soul}

\begin{document}

\newcommand{\ittext}[1]{\mbox{\rm\scriptsize #1}}
\hyphenation{to-mo-gra-phy}

\title{Coherent and incoherent excitation pathways in time-resolved photoemission orbital tomography of CuPc/Cu(001)-2O}

\author{A.~Adamkiewicz$^{1}$, M.~Raths$^{2,3,4}$, M.~Stettner$^{2,3,4}$, M.~Theilen$^{1}$, L.~Münster$^{1}$, S.~Wenzel$^{2,3}$, M.~Hutter$^{2,3,4}$, S.~Soubatch$^{2,3}$, C.~Kumpf$^{2,3,4}$, F.~C.~Bocquet$^{2,3}$, R.~Wallauer$^{1}$, F.~S.~Tautz$^{2,3,4,*}$, and U.~H{\"o}fer$^{1,5,*}$}
\address{$^1$Faculty of Physics and Material Sciences Center, Philipps University, 35032 Marburg, Germany\\
$^2$Peter Grünberg Institute (PGI-3), Jülich Research Centre, 52425 Jülich, Germany\\
$^3$Jülich Aachen Research Alliance (JARA), Fundamentals of Future Information Technology, 52425 Jülich, Germany\\
$^4$Institute for Experimental Physics IV-A, RWTH Aachen University, 52074 Aachen, Germany \\
$^5$Institute of Experimental and Applied Physics, University of Regensburg, 93040 Regensburg, Germany.  \\
$^*$E-mail: hoefer@physik.uni-marburg.de (U.H.); s.tautz@fz-juelich.de (F.S.T.)\looseness=-6 \\
}

\date{\today}

\begin{abstract}

{\sl {\bf Abstract:} 
Time-resolved photoemission orbital tomography (tr-POT) offers unique possibilities for tracing molecular electron dynamics. The recorded pump-induced changes of the angle-resolved photoemission intensities allow to characterize unoccupied molecular states in momentum space and to deduce the incoherent temporal evolution of their population. Here, we show for the example of CuPc/Cu(001)-2O that the method also gives access to the coherent regime and that different excitation pathways can be disentangled by a careful analysis of the time-dependent change of the photoemission momentum pattern. In particular, we demonstrate by varying photon energy and polarization of the pump light, how the incoherent temporal evolution of the LUMO distribution can be distinguished from coherent contributions of the projected HOMO. Moreover, we report the selective excitation of molecules with a specific orientation at normal incidence by aligning the electric field of the pump light along the molecular axis.\\}

\end{abstract}

\maketitle

\section{Introduction}

The concept of molecular orbitals plays a key role for understanding electronic and chemical properties of materials~\cite{Fukui52jcp,Woodward65jacs}. This has evoked considerable interest for direct imaging of frontier molecular orbitals, i.e., experimentally revealing their distribution either in real or reciprocal (momentum, $k$) space~\cite{Repp05prl1,Schmidt21}. Several years ago, the technique of photoemission orbital tomography (POT) was introduced and proved to be a powerful method to identify and even reconstruct orbitals of well-ordered molecular systems~\cite{Puschnig09sci,Stadtm12el,Wiessner14natcomm,Luftner14pnas,Weiss15natcomm,Offenb15jelsp,Yang19natcomm,Haags22scia}. This method relates the measured angular distribution of photoemission to the real space shape of specific orbitals; in the plane-wave approximation to the final state, one can employ the inverse Fourier transformation to relate the $k$ space orbital to the real space one~\cite{Gadzuk74aps,Puschnig09sci,Weiss15natcomm}. Recently, POT was successfully combined with a laser pump-probe scheme~\cite{Wallauer21sci,Neef23nat,Bennecke23arXiv}. In this way, POT offers access to unoccupied electronic states of molecules and can be used to investigate molecular charge transfer dynamics. 

In time-resolved POT (tr-POT), a short pump pulse electronically excites the molecule and a time-delayed probe pulse is used for photoemission. The time-dependent changes of the measured photoemission momentum distributions that result from such a two-photon photoemission (2PPE) experiment can have several origins. They can result from the population or depopulation of excited molecular states~\cite{Wallauer21sci}, from the formation of excitons~\cite{Neef23nat,Bennecke23arXiv}, or from nuclear motion. And on the attosecond time scale, where electronic wave packet imaging in molecular systems is of major interest~\cite{Nisoli17cr}, the momentum distribution is expected to reflect the temporal evolution of the electronic wavepacket launched by the pump pulse.

\begin{figure}[t!]
	\begin{center}
	\vspace{1.cm}
	\includegraphics[width=.9\columnwidth]{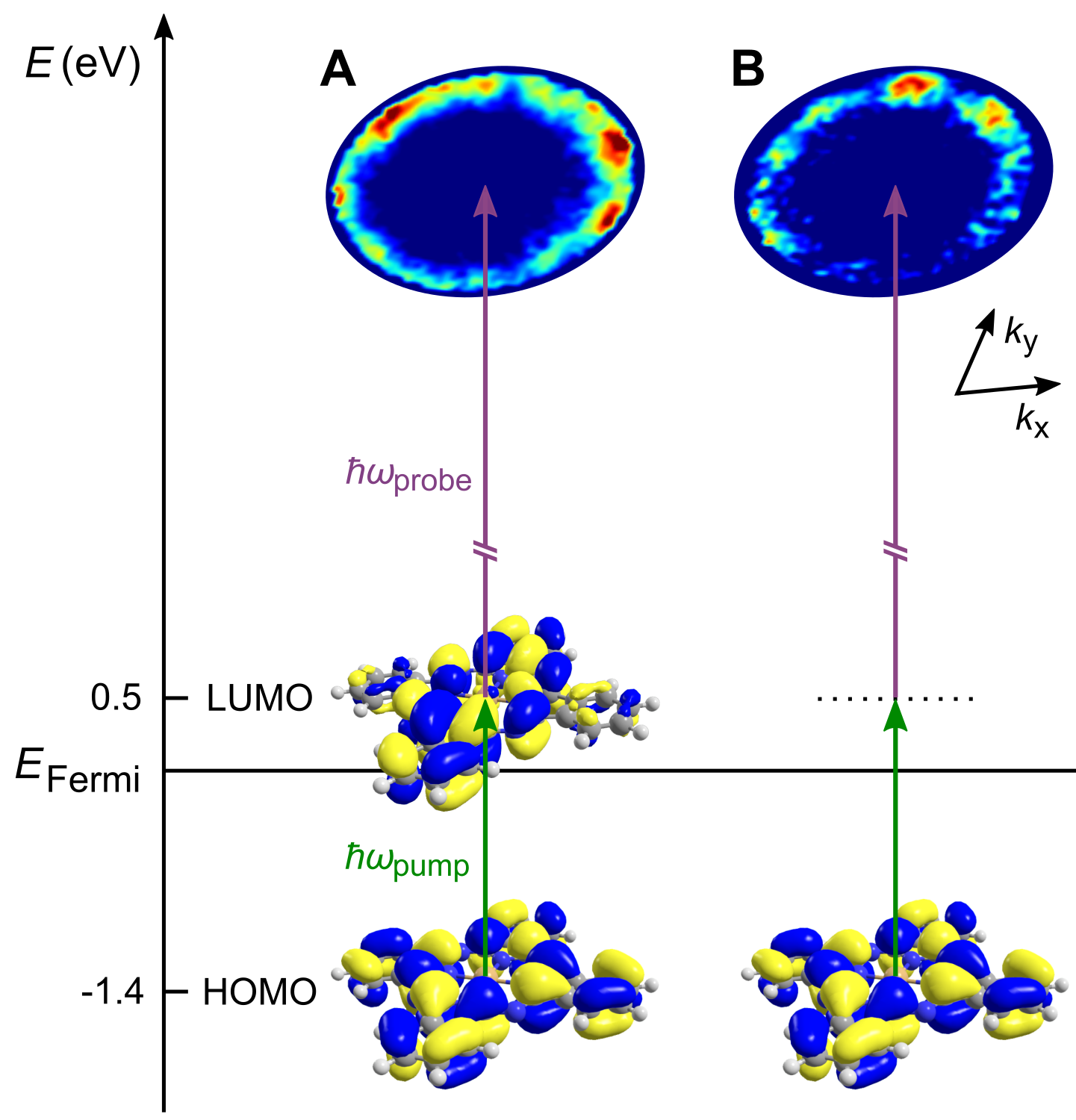}
	\vspace{0.2cm}
	\caption{Two-photon photoemission (2PPE) excitation scheme for emission from the CuPc LUMO and HOMO contributing to the measured 2PPE signal: emission from the transiently occupied CuPc LUMO populated via pump light \textbf{(A)} and direct two-photon emission from the HOMO \textbf{(B)}. The momentum distribution maps shown on top are measured on CuPc/Cu(001)-2O at delay time between pump and probe pulses $\Delta t$~= $-$11\,fs at 0.5\,eV, and with pump and probe photon energy $\hbar\omega_{\text{pump}}$~=~1.81\,eV and $\hbar\omega_{\text{probe}}$~=~21.7\,eV, respectively, for same (A) and crossed (B) polarization of pump and probe pulses.}
	\label{fig:exc_scheme}
	\end{center}
\end{figure}

For the evaluation of the time-dependent momentum distributions of photoemission it is important to consider that the signal from occupied states can significantly contribute to the measured 2PPE intensity during the temporal overlap of pump and probe pulses~\cite{Fauster95,Hertel96prl,Weinelt04prl,Armbrust12prl,Schubert12prb,Lerch18jp}. Using the example of copper(II) phthalocyanine (CuPc), Fig.~\ref{fig:exc_scheme} sketches the difference between the resonant transition from the highest occupied molecular orbital (HOMO) into the lowest unoccupied molecular orbital (LUMO) and subsequent excitation of the transient LUMO population on the one hand (Fig.~\ref{fig:exc_scheme}A) and the direct photoemission from the HOMO driven by a coherent two-photon excitation via a virtual intermediate state on the other (Fig.~\ref{fig:exc_scheme}B). The relative weights of both pathways depend on a number of factors~\cite{Klamroth01prb,Ueba00apa}. Besides the matrix elements of the 2PPE process, these are mainly the lifetime of the intermediate state (LUMO) population, the dephasing time of the induced HOMO-LUMO polarization, and the detuning of the pump photon energy from the HOMO$\rightarrow$LUMO transition energy.

Here, we show for the example of CuPc adsorbed on an oxygen-reconstructed Cu(001) surface how one can clearly disentangle coherent and incoherent excitation pathways by means of systematic tr-POT experiments. To this end, we tune the photon energy of the pump pulse across the HOMO-LUMO gap as well as vary its polarization. The measured kinetic energy of photoelectrons and their distribution in momentum space compare well with a model based on a density matrix calculation of the 2PPE process. The employed model exclusively accounts for a time-dependent population and depopulation of the excited LUMO, i.e., no temporal evolution of its emission pattern was assumed. For CuPc/Cu(001)-2O, the coherent 2PPE from the HOMO dominates the pronounced time-dependent changes in the momentum distribution pattern that are observed experimentally. In addition, we show that the excitations from the LUMO of molecules with specific orientations can be controlled by aligning the polarization of the pump light along the molecular diagonal axis of the CuPc.

\section{Experimental}

The experiments were performed in ultra-high vacuum (base pressure $<$~2~$\times$~10$^{-10}$\,mbar) on a well-ordered monolayer film of CuPc on a Cu(001)-$(\sqrt{2} \times 2\sqrt{2})R45^\circ$-2O missing-row-reconstructed surface.

The oxygen-induced reconstruction of Cu(001), which was prepared by exposure of a hot (430\,K) sputtered and annealed copper surface to O$_2$ (5~$\times$~10$^{-7}$\,mbar)~\cite{Moenig13acsn}, retains long-range order in the organic layer. At the same time, the molecules are expected to become decoupled from the metal~\cite{Yang18,Wallauer21sci}. The evaporation rate of CuPc was calibrated on Ag(110) using low-energy electron diffraction (LEED). After adsorption of CuPc on Cu(001)-2O, scanning tunneling microscopy (STM) and LEED confirmed a long-range-ordered molecular overlayer with one flat-lying CuPc molecule per quadratic unit cell of 210\,$\text{\AA}^{2}$. Molecules are oriented with their long diagonal axes $\pm$25° away from the [010] direction of the Cu crystal, forming two symmetry-equivalent mirror domains (cf. Supporting Information (SI), section~1).

The distributions of the photoemitted electrons were acquired using a time of flight momentum microscope~\cite{Wallauer21sci}. In a femtosecond 2PPE scheme, tunable pump pulses in the visible spectral range (pulse duration $\tau_{\text{pump}}$~= 40 - 85\,fs) were combined with extreme ultraviolet probe pulses of 21.7\,eV (pulse duration $\tau_{\text{probe}}$~= 20\,fs), which were generated by high laser harmonics~\cite{Heyl12jp,Wallauer21sci}. Close to normal and inclined (70° with respect to surface normal) incidence geometries were used for the pump beam; they will be further referred to as s- and p-polarization, respectively. The angle of incidence of the p-polarized probe beam was fixed at 70°. To realize normal incidence of the pump light, we implemented a mirror inside the momentum microscope lens system.

The deposition of CuPc as well as all LEED and photoelectron spectroscopy measurements were performed at room temperature, the STM measurements at 6.2\,K.

In this work, all calculated momentum distribution maps were created following Ref.~\cite{Brands21}. This routine assumes photoemission from isolated molecules in the gas phase. A comparison to our results is justified, considering that in our tr-POT experiments CuPc is only weakly coupled to the Cu(001)-2O substrate.

\section{Results and Discussion}

\begin{figure}[t!]
	\begin{center}
	\vspace{0.cm}
	\includegraphics[width=1\columnwidth]{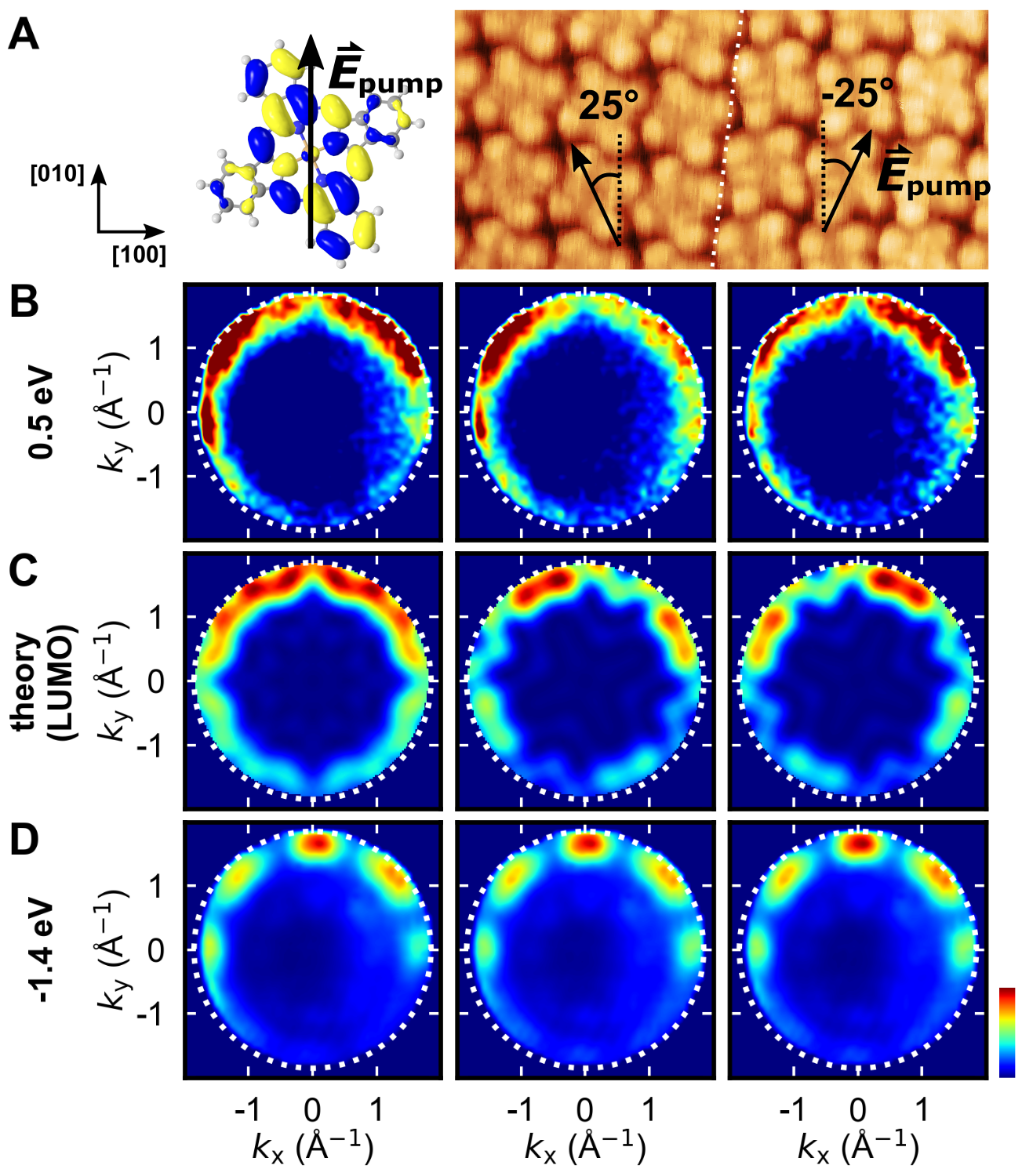}
	\vspace{-.3cm}
	\caption{\textbf{(A)} Left: CuPc LUMO (one state of $e_{\text{g}}$ representation). Right: STM image (9~x 4\,nm$^{\text{2}}$, $U_{\text{t}}$~= $-$0.6\,V, $I_{\text{t}}$~= 1\,nA, $T_{\text{sample}}$~= 6.2\,K) of a CuPc monolayer showing a border (white dotted line) between two symmetry-equivalent mirror domains. The arrows indicate the orientation of the pump pulse electric field vector $\vec{\mathbf{E}}_{\mathbf{pump}}$ which is oriented within 0°, 25° and $-$25° with respect to [010] high symmetry direction of the Cu(001) crystal for the left, center and right columns of the tomograms (B)-(D), respectively. \textbf{(B)} Momentum distribution maps measured at the energy corresponding to the emission from the CuPc LUMO (0.5\,eV). \textbf{(C)} Tomograms simulated for the LUMO of two CuPc molecules rotated $\pm$25° with respect to [010]. The left panel shows a superposition of both maps with equal weight, the center and right panel show only contributions from 25° and $-$25°, respectively. \textbf{(D)} $k$ maps measured at the energy corresponding to the 2PPE emission from the CuPc HOMO ($-$1.4\,eV). $k_{\text{x}}$ and $k_{\text{y}}$ match with [100]* and [010]* direction, respectively. (B) and (D) are measured at delay time $\Delta t$~= 18\,fs.}
	\label{fig:0deg}
	\end{center}
\end{figure}

We start with demonstrating a preferential excitation of CuPc molecules from a particular mirror domain by changing the orientation of the electric field of the s-polarized pump pulse. To this end, we set the photon energy of the pump light to 1.91\,eV~\cite{Edwards70} in order to match the HOMO$\rightarrow$LUMO $S_1$ transition of adsorbed CuPc and thus to obtain a resonant excitation of molecules. Note that with normal incidence of the pump beam, the electric field vector of the latter is parallel to the surface and the molecular plane. Excitations induced by the pump pulse are therefore solely driven by in-plane polarization. At first, we align the polarization of the pump light, i.e., its electric field vector, along the [010] direction of copper (cf. Fig.~\ref{fig:0deg}A) and measure the momentum distribution of photoelectron emission induced by the 21.7\,eV probe photons. The momentum distribution map ($k$ map) in Fig.~\ref{fig:0deg}B (left panel) is measured at the kinetic energy of photoelectrons corresponding to emission from the transiently occupied CuPc LUMO with its orbital energy $\epsilon_{\text{LUMO}}$~= 0.5\,eV. Note that the energies of electronic states are defined with respect to the Fermi level $E_{\text{F}}$ (fixed as $E$~= 0). We compare this experimental $k$ map with the simulated tomogram. For the simulation, we calculated the LUMO pattern for two symmetry-equivalent domains oriented $\pm$25° with respect to [010] and assumed both domains to be contributing equally to the simulated momentum distribution (in agreement with STM observations, cf. SI, section~1). Comparing the measured (Fig.~\ref{fig:0deg}B left panel) and simulated (Fig.~\ref{fig:0deg}C left panel) $k$ maps, we see a good resemblance. The momentum distribution measured at the kinetic energy corresponding to direct excitation from the CuPc HOMO (orbital energy $\epsilon_{\text{HOMO}}$~= $-$1.4\,eV, i.e., below Fermi level, cf. Fig.~\ref{fig:exc_scheme}) reveals a very different intensity pattern (Fig.~\ref{fig:0deg}D left panel). 

Now we rotate the pump pulse polarization by 25°, such that the electric field vector becomes aligned with either of two orientations of CuPc present at the surface (cf.~Fig.~\ref{fig:0deg}A, STM image). This significantly affects the measured $k$ maps of the LUMO (Fig.~\ref{fig:0deg}B center and right panels): The intensity distribution becomes clearly asymmetric with respect to the [010]* direction. This is in turn confirmed by theoretical simulations (Fig.~\ref{fig:0deg}C center and right panels). Because both experimental and theoretical $k$ maps are results of photoemission from two ensembles of differently oriented molecules, it is natural to assign the observed asymmetry to a different contribution of the two ensembles depending on their alignment relative to the electric field of the pump light. The agreement between the experimental data and the calculated $k$ maps at the chosen orientations of the electric field clearly confirms that the excitations occur within the molecules due to the HOMO$\rightarrow$LUMO transition. Note that the CuPc LUMO level is a degenerate combination of two states of $e_{\text{g}}$ representation concentrated along two perpendicular axes of the molecule connecting its opposite isoindole entities, while the $a_{\text{1u}}$ HOMO spreads over the entire molecular backbone~\cite{Luftner14jesrp}, cf. Fig.~\ref{fig:exc_scheme}. Hence, the excitation due to HOMO$\rightarrow$LUMO transition must be maximized when the electric field of the resonant pumping is parallel to one of the molecular axes. Furthermore, we find that the orientation of the electric field of the pump beam does not affect the $k$ maps of the HOMO caused by direct emission from this orbital (Fig.~\ref{fig:0deg}D) which is solely driven by the probe pulses. This also means that the electric field of our pump pulses is not strong enough to induce visible changes in the HOMO population.

\begin{figure*}[t!]
	\begin{center}
	\includegraphics[width=1.6\columnwidth]{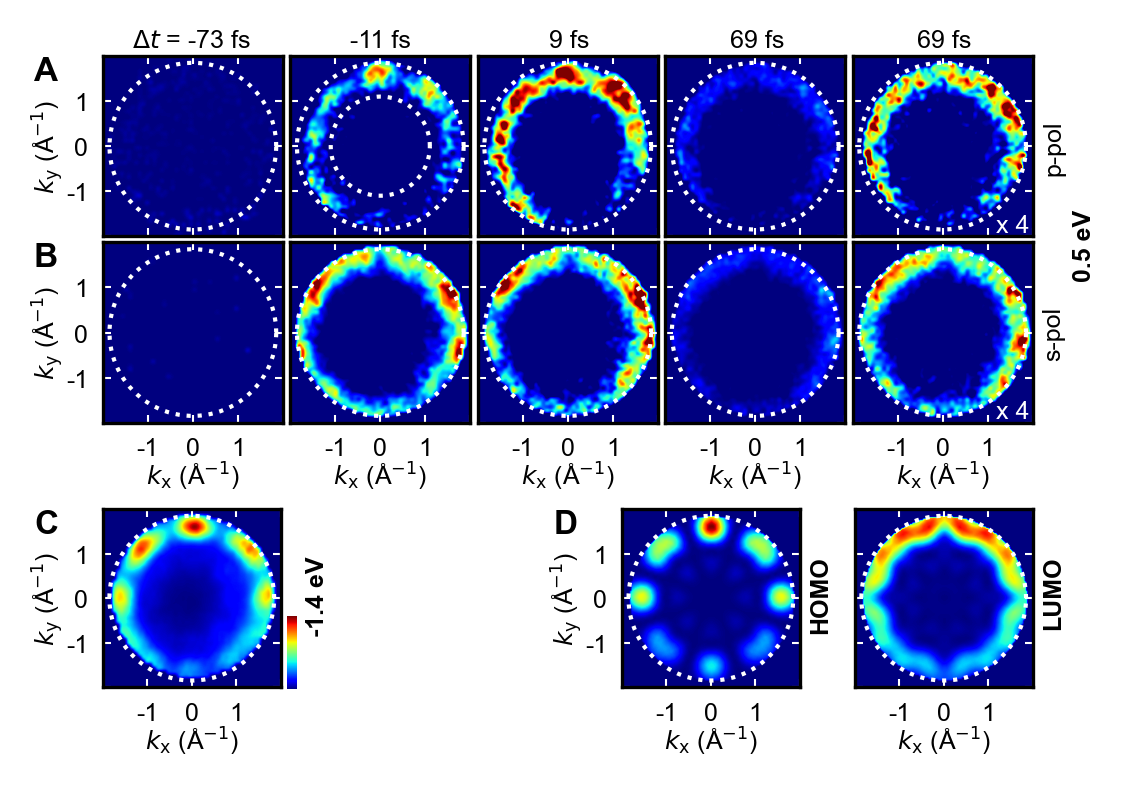}
	\vspace{-0.3cm}
	\caption{Time-dependent experimental $k$ maps of unoccupied states ($E$~= 0.5\,eV) for \textbf{(A)} p-polarized (70° incidence) and \textbf{(B)} s-polarized (0° incidence) pump pulses ($\hbar\omega_{\text{pump}}$~= 1.81\,eV) at different delay times $\Delta t$ between pump and probe pulses. \textbf{(C)} Experimental $k$ map of the CuPc HOMO ($E$~= $-$1.4\,eV). \textbf{(D)} Theoretical $k$ maps for CuPc HOMO and LUMO, calculated assuming equal contributions from $-$25° and $+$25° oriented molecules. Intensity in (A) and (B) is normalized in reference to the signal of the Cu d band for all measurements, respectively.}
	\label{fig:kmaps_delay} 
	\end{center}
\end{figure*}

Having established the effect of the orientation of the electric field in the s-polarized pump light on the resonant excitation of the LUMO, we now proceed to disentangle different excitation pathways which occur in the applied pump probe scheme. The separation and identification of different pathways is a well-known, nevertheless sometimes challenging issue in 2PPE experiments~\cite{Fauster95,Hertel96prl,Tegeder07apa,Armbrust12prl,Schubert12prb,Ilyas14aps,Bogner15pccp,Yamada18pss,Lerch18jp}. 
One possibility to distinguish the 2PPE intensity of occupied states from that of unoccupied states is to tune the photon energy of pump and probe pulses. The kinetic energy of the photoemitted electrons is expected to shift with the sum $\hbar\omega_{\text{pump}}+\hbar\omega_{\text{probe}}$ if they originate from occupied states whereas unoccupied states should exhibit a shift with $\hbar\omega_{\text{probe}}$ only \cite{Fauster95,Armbrust12prl}. In time-resolved 2PPE, unoccupied states are characterized by a finite lifetime, whereas the signal from occupied states ideally follows the cross-correlation between pump and probe pulses~\cite{Hertel96prl}.

In cases where the presence of several unoccupied states or a very short lifetime makes an unambiguous assignment difficult, a measurement of the dispersion of the involved states with angle-resolved 2PPE is extremely helpful, e.g., for surface states of metals~\cite{Schubert12prb}. 
When one applies 2PPE to organic molecular systems with many spatially localized states~\cite{Tegeder07apa,Ilyas14aps,Bogner15pccp,Yamada18pss,Lerch18jp} one often faces the problem that indeed many, and sometimes broad, states overlap but that they cannot be separated by their dispersion. 
The present results show that tr-POT can be of exceptional advantage for disentangling different contributions to the photoemission signal by making use of characteristic momentum fingerprints of molecular orbitals.

As first means to disentangle the different 2PPE pathways, we use tr-POT to reveal the dynamics of the photoemission intensity using a controlled delay time $\Delta t$ between pump and probe pulses. Fig.~\ref{fig:kmaps_delay}A shows $k$ maps measured at different $\Delta t$ at 0.5\,eV above the Fermi edge using p-polarized pump light (angle of incidence 70°) with photon energy $\hbar\omega_{\text{pump}}$~= 1.81\,eV. 
The change of the momentum distribution pattern with the delay time is obvious: at $\Delta t$~= $-$11\,fs, the intensity distribution shows strong resemblance to the measured (Fig.~\ref{fig:kmaps_delay}C) and calculated (Fig.~\ref{fig:kmaps_delay}D left) intensity patterns of the CuPc HOMO. Then, with increasing delay time, the pattern gradually evolves until, at $\Delta t$~= 69\,fs, it evidently resembles the LUMO momentum fingerprint (compare Fig.~\ref{fig:kmaps_delay}A and \ref{fig:kmaps_delay}D right). For the calculated tomograms in (D), an equal ratio of both molecular orientations as discussed above was assumed. From the evolution of the $k$ maps with delay time in Fig.~\ref{fig:kmaps_delay}A, one can directly derive that the signal at 0.5\,eV contains (at least) two contributions closely related to the CuPc HOMO and LUMO, with noticeably different dynamics.

When applying p-polarized pump light in combination with the p-polarized probe beam, excitation based on coherent phenomena is enabled. Therefore, during temporal overlap between the two beams, emission from the naturally occupied electronic states driven by coherent two-photon excitation may contribute to the observed signal. Taking the characteristic HOMO momentum pattern and kinetic energy of the photoemission into account, we find that the excitation pathway dominating during temporal overlap between pump and probe pulses in Fig.~\ref{fig:kmaps_delay}A is direct two-photon photoemission from the HOMO. After temporal overlap ($\Delta t\gtrsim$50\,fs), the signal is dominated by emission from the CuPc LUMO: electrons are first excited into the CuPc LUMO via pump light (HOMO$\rightarrow$LUMO transition). They experience a finite lifetime in this intermediate electronic state and can be probed with the second beam even after temporal overlap. Note that only the significant differences in the $k$ fingerprints of CuPc LUMO and HOMO allow for clear identification of the involved orbitals in the time-dependent $k$ maps shown in Fig.~\ref{fig:kmaps_delay}A. If this condition is met, tr-POT is capable of disentangling 2PPE pathways of different dynamics in $k$ space, even for different contributions with similar kinetic energies of the photoemitted electrons.

If we change the incidence angle of the pump light from 70° to close to 0°, i.e., turning its polarization from p to s, the probability of the coherent two-photon excitation is strongly suppressed due to the large angle between the electric fields in the two beams. This is illustrated in Fig.~\ref{fig:kmaps_delay}B: evidently, only emission from the LUMO contributes to the signal and no temporal evolution of the momentum distribution pattern is observed. Comparing the results of the two experiments with different polarization of the pump light, it appears that p-polarized pulses (Fig.~\ref{fig:kmaps_delay}A) clearly activate an entirely different excitation pathway $-$ a direct two-photon emission from the HOMO.

\begin{figure}[th!]
	\begin{center}
	\includegraphics[width=\columnwidth]{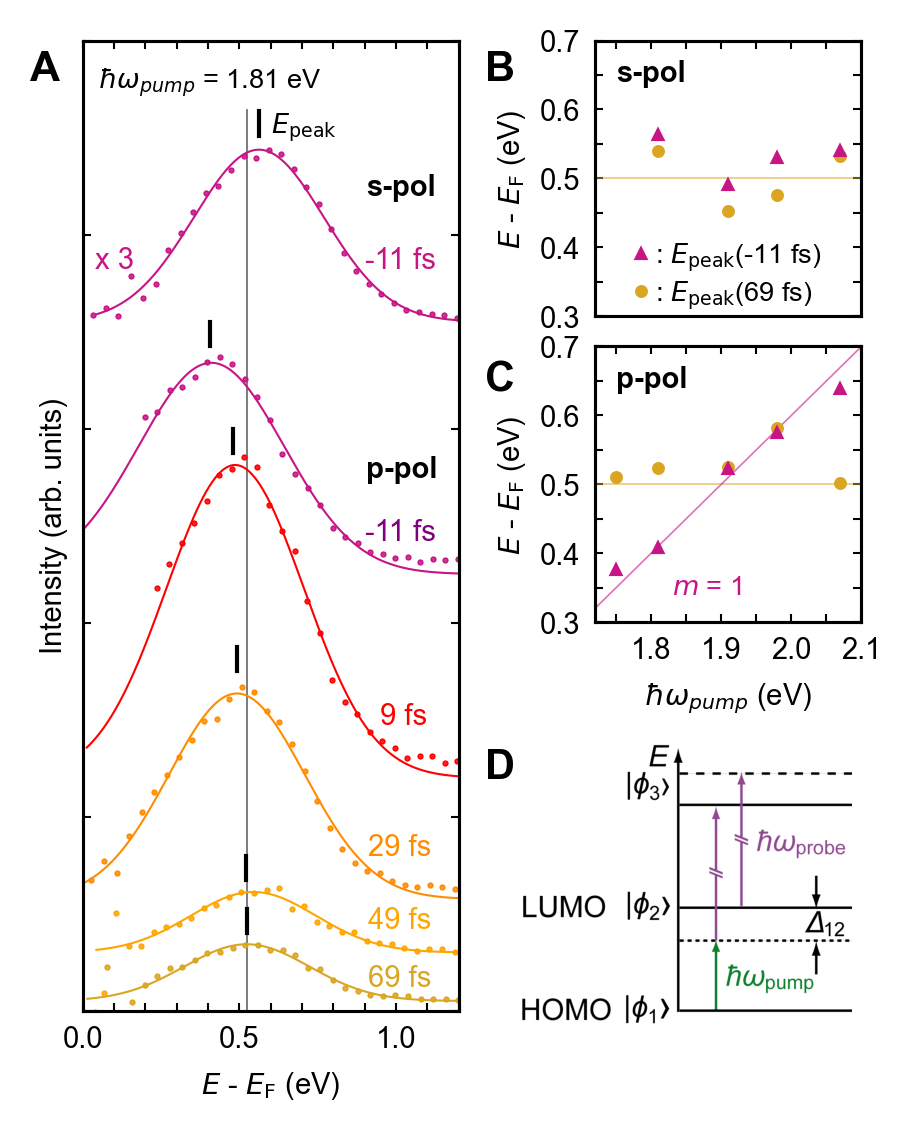}
	\vspace{-0.5cm}
	\caption{\textbf{(A)} Energy spectra around 0.5\,eV above the Fermi level for pump photon energy $\hbar\omega_{\text{pump}}$~= 1.81\,eV at different delay times $\Delta t$ and for different pump polarizations, integrated within boundaries in $k_x$ and $k_y$ as indicated by the white dotted lines in Fig.~\ref{fig:kmaps_delay}A ($\Delta t$~= $-$11\,fs). A background following the Fermi function was subtracted (s.~SI, section~2). Dots represent experimental data, the lines show Voigt fits to the data. Intensity in (A) is normalized in reference to the Cu d band signal. The energy of the respective peak maximum for early ($\Delta t$~= $-$11\,fs) and late ($\Delta t$~= 69\,fs) delay times is plotted for different $\hbar\omega_{\text{pump}}$ for \textbf{(B)} s-polarized and \textbf{(C)} p-polarized pump light. \textbf{(D)} Energy scheme for coherent two-photon excitation from the HOMO and excitation from the transiently populated LUMO. $\Delta_{\text{12}}$ indicates a positive detuning of the pump photon energy with respect to the HOMO$\rightarrow$LUMO $S_1$ transition.}
	\label{fig:E_spectra} 
	\end{center}
\end{figure}

\begin{figure*}[t!]
	\begin{center}
	\includegraphics[width=1.6\columnwidth]{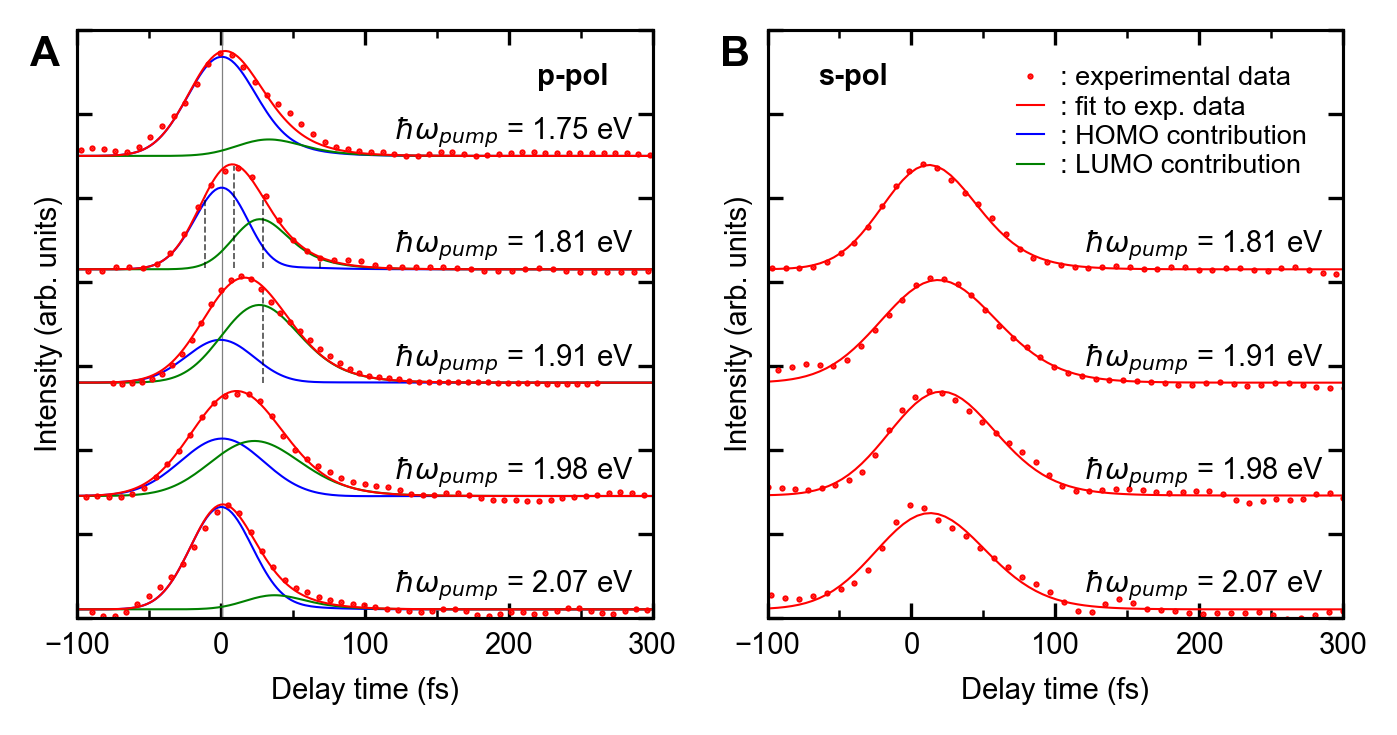}
	\vspace{-0.1cm}
	\caption{Temporal evolution of the photoemission intensity for \textbf{(A)} p-polarized and \textbf{(B)} s-polarized pump light for different pump photon energies. For the measured delay scans, the data was integrated over 0.34 - 0.98\,eV within boundaries in $k_x$ and $k_y$ as indicated by the white dotted lines in Fig.~\ref{fig:kmaps_delay}A ($\Delta t$~= $-$11\,fs) in order to minimize background noise. The fit is based on solving the optical Bloch equations for a three-level system (more details discussed in the text and in SI, section~3). For (A), the HOMO and LUMO contributions extracted from the fits are shown. For (B), no photoemission from the projected HOMO was observed. The dotted vertical lines in (A), $\hbar\omega_{\text{pump}}$~= 1.81\,eV and 1.91\,eV, show designated delay times relating to $k$ maps shown in Fig.~\ref{fig:kmaps_delay} and Fig.~\ref{fig:kmaps_deconvolution}.}
	\label{fig:delay_scans} 
	\end{center}
\end{figure*}

In order to energetically disentangle and further confirm the origin of both contributions, we systematically tune the pump photon energy across the HOMO-LUMO gap. Fig.~\ref{fig:E_spectra}A shows the time-dependent photoemission spectra of the data shown in Fig.~\ref{fig:kmaps_delay}A,B. The pump photon energy $\hbar\omega_{\text{pump}}$~= 1.81\,eV corresponds to slightly off-resonant excitation ($\Delta_{\text{12}}$~= 135\,meV, $\Delta_{\text{12}}$: detuning of the pump photon energy with respect to the HOMO$\rightarrow$LUMO $S_1$ transition, cf.~Fig.~\ref{fig:E_spectra}D). A background following the Fermi function was subtracted in order to clearly separate the signal around 0.5\,eV from bulk emission at the Fermi level (cf. SI, section~2). For s-polarized pump light, the spectrum peaks at the energy $E_{\text{peak}}$ close to the LUMO level, $\epsilon_{\text{LUMO}}$~= 0.5\,eV, and does not shift with delay time (the example of $\Delta t$~= $-$11\,fs is shown in Fig.~\ref{fig:E_spectra}A (purple curve)). In contrast, for p-polarized pump light, the spectra show a gradual shift with increasing delay according to the amount of detuning $\Delta_{\text{12}}$. While at early delay ($\Delta t$~= $-$11\,fs) the emission from direct HOMO excitation dominates the signal, for late delay ($\Delta t\gtrsim$50\,fs), mostly emission from the LUMO is contributing. Therefore, the energy of HOMO and LUMO contributions can be determined at early and late delay times, respectively. 

Fig.~\ref{fig:E_spectra}B shows the energy of the peak maxima for early ($\Delta t$~= $-$11\,fs) and late ($\Delta t$~= 69\,fs) delay times for different pump photon energies. Evidently, no shift is observed for s-polarized pump light. The data is scattered around the energy of the LUMO, $\epsilon_{\text{LUMO}}$~= 0.5\,eV, for both delay times and different pump photon energies. For p-polarized pump light (Fig.~\ref{fig:E_spectra}C), systematic tuning of the pump photon energy allows the unambiguous separation of both excitation pathways. Comparison of the energy of the peak maxima at both early and late delay times reveals the different dependences of both contributions on the pump photon energy (Fig.~\ref{fig:E_spectra}D): early delay signals are dominated by the projected HOMO contribution and are therefore of linear dependence on the pump photon energy. In contrast, resonant LUMO excitation is the main contribution to the signal at late delay times. The peak maxima of the late delay spectra show no dependence on pump photon energy and are consistent with the energy of the CuPc LUMO.

For a quantitative analysis, we performed density matrix calculations and solved the optical Bloch equations for a three-level system. Time-dependent intensity traces were fitted to the experimental data for different pump photon energies for p- and s-polarized pump pulses (Fig.~\ref{fig:delay_scans}). The temporal evolution of the signal around 0.5\,eV was obtained by integrating over the energy range 0.34 - 0.98\,eV within the momentum integration range indicated by the two dotted circles in Fig.~\ref{fig:kmaps_delay}A, $\Delta t$~= $-$11\,fs. The common fit parameters are consistent for all delay curves, while the detuning $\Delta_{\text{12}}$ was changed according to the applied pump photon energy. For p-polarized pump light (Fig.~\ref{fig:delay_scans}A), the transients broken down into HOMO and LUMO contributions are shown. Closest to resonant excitation, $\hbar\omega_{\text{pump}}$~= 1.91\,eV, LUMO population is most efficient, thus being a major contribution to the measured signal. For $\hbar\omega_{\text{pump}}$~= 1.75\,eV and $\hbar\omega_{\text{pump}}$~= 2.07\,eV, the intensity traces mostly consist of the contribution from the projected HOMO. For the fits of the data measured with s-polarized pump light (Fig.~\ref{fig:delay_scans}B), coherent contributions were not taken into account. A more detailed discussion on the calculations can be found in the SI, section~3. The lifetime of the LUMO population was determined to be $\tau$~= (20$\pm$5)\,fs. The coupling of CuPc to the substrate is stronger than for PTCDA, where $\tau$ was found to be $\simeq$~250\,fs~\cite{Wallauer21sci}.

\begin{figure*}[t!]
	\begin{center}
	\includegraphics[width=1.85\columnwidth]{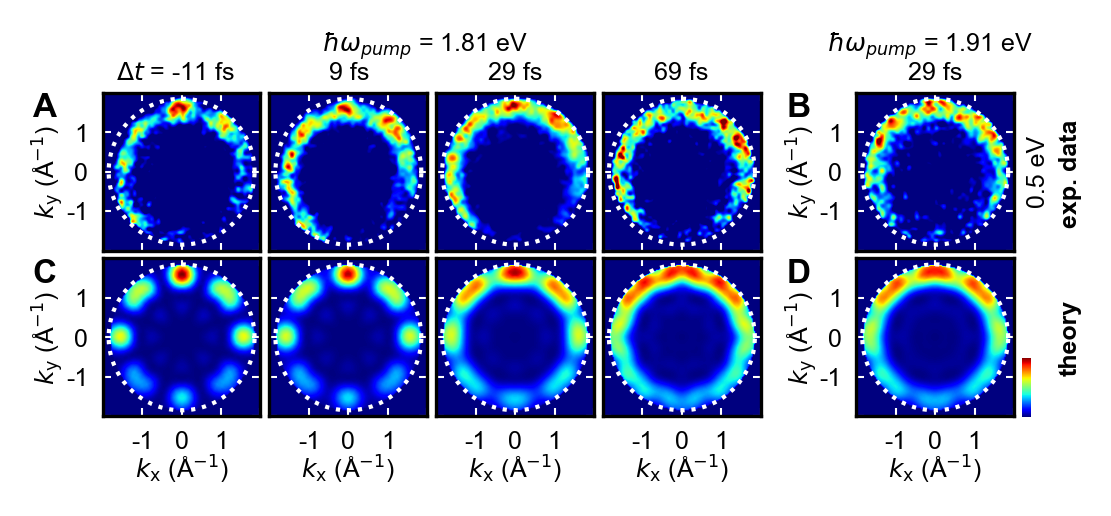}
	\vspace{-.3cm}
	\caption{Measured $k$ maps for p-polarized pump light with $\hbar\omega_{\text{pump}}$~= 1.81\,eV \textbf{(A)} and $\hbar\omega_{\text{pump}}$~= 1.91\,eV \textbf{(B)} at different delay times $\Delta t$. \textbf{(C), (D)} Calculated tomograms: Superposition of the theoretical HOMO and LUMO momentum distributions (cf.~Fig.~\ref{fig:kmaps_delay}D). The relative weights of the contributions was extracted from the fit result shown in Fig.~\ref{fig:delay_scans}A. The dashed vertical lines in Fig.~\ref{fig:delay_scans}A indicate the designated delay times at which the momentum maps in (A) and (B) were obtained.}
	\label{fig:kmaps_deconvolution} 
	\end{center}
\end{figure*}

Knowing the temporal evolution of the relative weight of HOMO and LUMO contributions to the 2PPE signal from the density matrix calculations performed for Fig.~\ref{fig:delay_scans}A, we simulated time-dependent $k$ maps for p-polarized pump light with $\hbar\omega_{\text{pump}}$~= 1.81\,eV and 1.91\,eV. Fig.~\ref{fig:kmaps_deconvolution} shows a comparison of tomograms measured at 0.5\,eV (A), (B) and simulated $k$ maps (C), (D) for different delay times. The $k$ maps in (C) and (D) are a superposition of theoretical HOMO and LUMO momentum distributions of CuPc for two symmetry-equivalent mirror domains, oriented $\pm$25° with respect to the [010] orientation of the copper surface. The relative weights of HOMO and LUMO contributions at the designated delay times $\Delta t$ were taken from the fit result shown in \ref{fig:delay_scans}A. The measured (A), (B) and simulated (C), (D) $k$ maps show remarkable resemblance for all delay times. Even more, comparing the data for $\Delta t$~= 29\,fs for both pump photon energies, the different relative weights of HOMO and LUMO contributions due to different detuning of the pump photon energy with respect to resonant HOMO$\rightarrow$LUMO transition, $\Delta_{12}$, are visible in both measured and calculated $k$ maps. From the density matrix calculations, we derived the ratio 1 : 2.1 (1 : 4.0) for $\hbar\omega_{\text{pump}}$~= 1.81\,eV ($\hbar\omega_{\text{pump}}$~= 1.91\,eV) of HOMO and LUMO contributions, respectively (Fig.~\ref{fig:delay_scans}A). In both measured and simulated tomograms at $\Delta t$~= 29\,fs, the characteristic HOMO intensity peaks are more pronounced for $\hbar\omega_{\text{pump}}$~= 1.81\,eV ($\Delta_{\text{12}}$~= 135\,meV) than for $\hbar\omega_{\text{pump}}$~= 1.91\,eV ($\Delta_{\text{12}}$~= 35\,meV). This is caused by less efficient population of the CuPc LUMO via the pump light for larger detuning ($\Delta_{\text{12}}$~= 135\,meV), resulting in lower intensity of emission from the LUMO contributing to the $k$ maps. Overall, the intensity distributions in Fig.~\ref{fig:kmaps_deconvolution} reveal how 2PPE pathways with different dynamics can be deconvolved in momentum space via time-dependent POT measurements and simulations based on density matrix calculations, even accounting for different detuning of the pump photon energy with respect to resonant excitation.

\section{Conclusions}

We showed that time-resolved photoemission orbital tomography is an extremely useful method to separate coherent and incoherent excitation pathways in well-ordered molecular systems. In order to disentangle the HOMO and LUMO excitation contributions, we take the well-established 2PPE time- and pump photon energy-dependent routine a step further by measuring the momentum distribution via orbital tomography. This allows us to trace momentum orbital fingerprints in a time-resolved fashion. Even more, tr-POT is also clearly suited to reveal transient excitonic signatures due to time-dependent pattern changes visible in the momentum maps and will therefore be a compellingly useful tool when applied to other molecular systems, especially organic heterostructures.

\section{Acknowledgement}
The authors acknowledge financial support from the Deutsche Forschungsgemeinschaft (DFG) through SFB~1083, project-ID 223848855, and from the European Research Council (ERC) Synergy Grant, project-ID 101071259.\\

\textbf{Author contributions:} F.S.T. and U.H. conceived the research. M.R., M.S., and F.C.B. prepared the samples. M.R., M.S., S.W., M.H., S.S., C.K., F.C.B., and F.S.T. characterized the samples with LEED and STM. A.A., M.R., M.S., M.T., and L.M. performed the tr-POT experiments. A.A., R.W., and U.H. analyzed the tr-POT data. A.A., S.S., and U.H. wrote the paper with contributions from all other authors.

\textbf{Competing interests:} The authors have no competing interests. 

\textbf{Supporting Information:} Sample characterization, background subtraction of energy spectra, density matrix calculations for the three-level system.

\newpage

\bibliography{Adamkiewicz_CuPc_Cu001O2_v5c}
\end{document}